# Mechanistic insights into hydrogen reduction of multicomponent oxides via *in-situ* high-energy X-ray diffraction

Shiv Shankar[1,*], Barak Ratzker[1], Claudio Pistidda[2], Dierk Raabe[1], Yan Ma[1,3,*]

[1] Max Planck Institute for Sustainable Materials, Max-Planck-Str. 1, 40237, Düsseldorf, Germany

[2] Institute of Hydrogen Technology, Helmholtz-Zentrum hereon GmbH, D-21502 Geesthacht, Germany

[3] Department of Materials Science & Engineering, Delft University of Technology, Mekelweg 2, 2628 CD, Delft, the Netherlands

*Corresponding authors: s.shankar@mpi-susmat.de (S. Shankar), yan.ma@tudelft.nl (Y. Ma)

**Abstract**

Co-reduction of multicomponent oxides with hydrogen provides a carbon-neutral approach toward sustainable alloy design. Herein, we investigate the hydrogen-based direct reduction, using *in-situ* high-energy X-ray diffraction of two precursor variants, including, mechanically mixed powders and pre-sintered oxide mixtures, targeting an equiatomic CoFeMnNi alloy. We find distinct reduction pathways and microstructure evolution depending on initial precursors. Mixed powders at 700 ºC are reduced to body-centered-cubic, face-centered-cubic, and MnO phases via halite, spinel, and $Mn_3O_4$ intermediates, whereas the pre-sintered material directly transforms into a mixture of metallic and oxide phases. The post-reduction microstructures are also different: mixed oxides show loosely packed morphology, whereas pre-sintered material reveals metallic nanoparticles supported on nanoporous MnO. The formation of nanoporous metallic networks is strongly governed by the precursor state, highlighting the role of initial precursors on the final microstructure. This precursor design strategy offers a single-step route to nanoporous alloys with potential applications in catalysis, and energy technologies.

Decarbonization of the metallurgical industry requires the development of sustainable routes for metal production, since current processes remain heavily reliant on fossil-based reductants [1,2]. Traditional alloy synthesis is not only energy intensive, involving multiple high-temperature processes such as melting, homogenization, and casting, but also poses substantial strain on the environment as it contributes to nearly 40% of all industrial greenhouse gas emissions [3].

Co-reduction of multicomponent oxides with hydrogen has therefore emerged as a sustainable approach to directly synthesize alloys with targeted applications [4]. Various studies have investigated the hydrogen reduction behavior of mixed oxides, including, binary [5–9], ternary [10,11], and multicomponent oxide systems [12–14]. With increasing chemical complexity, the reduction mechanisms become more complex due to the different thermodynamic stabilities of the constituent metal oxides [15]. In our previous study, we demonstrated that the initial precursor state, whether mechanically mixed powders or pre-sintered solid solution oxide mixtures, plays a decisive role in governing both the reduction pathway and the final microstructure [15]. This is illustrated by the fact that, although both precursor types reached similar reduction degrees (~80%) and primarily formed face-centered-cubic (FCC) and MnO phases, the pre-sintered sample uniquely exhibited an additional ~1 wt.% body-centered-cubic (BCC) phase. This is attributed to the localized deficiencies of FCC-stabilizing elements (Co and Ni) for Fe partitioned out of the (Fe,Mn)O solid solution. These results underscored that decoding the multiple reduction steps, particularly the formation of metastable intermediate phases, is critical for understanding hydrogen-based direct reduction (HyDR) mechanisms of multicomponent oxides.

Beyond precursor effects, co-reduction is also influenced by HyDR processing parameters, such as hydrogen partial pressure [16,17], temperature [18], and heating rate [19]. Despite advances in understanding the reduction behavior of mixed oxides, most studies [20–22] relied on thermogravimetric analysis and post-mortem microstructure characterization, failing to capture the real-time transient intermediate phases governing the underlying reduction mechanisms and microstructure evolution.

Therefore, we studied the HyDR process using *in-situ* synchrotron high-energy X-ray diffraction (HEXRD) of two precursor oxide mixtures: mechanically mixed powders and chemically mixed pre-sintered oxide mixtures. The *in-situ* reduction measurements provided mechanistic insights into multicomponent oxide reduction by monitoring the reduction sequence and the transient phases involved in the reduction process. Microstructure characterization revealed distinctly different behavior, mixed powder reduced to a loosely packed metallic and oxide phase, whereas the pre-sintered sample developed a dual phase microstructure, comprising metallic nanoparticles supported on a nanoporous manganese oxide matrix.

To obtain the oxide precursors, four metal oxide powders, namely – $Co_3O_4$, $Fe_2O_3$, $Mn_2O_3$, and NiO, were mixed with a targeted equiatomic metallic concentration (25 at.% each). The powders were mixed and homogenized using a planetary ball mill (Fritsch 7). Some of the mixed powder samples were also subsequently compacted and sintered in an Ar atmosphere at 1100 °C to obtain an atomically blended pre-sintered sample consisting of Co,Ni-rich halite and Mn,Fe-rich spinel. Detailed sample preparation procedure and characterization of initial precursors can be found elsewhere [15]. A rectangular prism specimen (~0.5×0.5×2 $mm^3$) was cut from the sintered sample using a diamond wire saw to fit into a sapphire capillary (0.6 mm inner diameter) used for the *in-situ* HEXRD measurement. The mixed powder was manually compressed into a green body in a capillary with the help of copper wires from both ends.

The *in-situ* synchrotron HEXRD reduction experiments were performed at 700 °C in hydrogen atmosphere using a capillary cell [23], as illustrated schematically in **Figure 1**. The measurements were conducted at the Powder Diffraction and Total Scattering Beamline, P02.1 of PETRA III in the Deutsches Elektronen-Synchrotron (DESY) [24], using 60 keV (λ=0.207381 Å) X-rays. Samples were placed between the incident beam and a Varex XRpad 4343CT fast area detector (2880×2880 pixels) with a sample-to-detector distance of ~1700 mm and a beam size of 0.5×0.5 $mm^2$. Debye-Scherrer diffraction rings were continuously recorded with an exposure time of 5 s. The capillary cell was heated by a ceramic resistive heater located beneath the capillary and the sample temperature was measured by a type K thermocouple. After flushing the cell with Ar for 5 min, $H_2$ (99.999%) was introduced at 2 bar total pressure from the gas inlet. The samples were heated to 700 °C with a ramping rate of 10 °C/min, held isothermally at 700 °C for 30 min, and then cooled down to room temperature within 15 min.

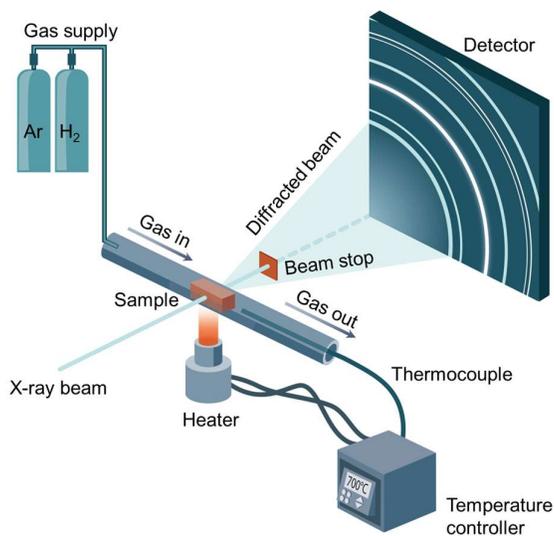

**Figure 1**. Schematic illustration of the *in-situ* synchrotron high-energy X-ray diffraction (HEXRD) experimental setup coupled with a capillary cell and a ceramic resistive heater to investigate the reduction behavior of multicomponent oxides in hydrogen atmosphere.

Diffraction patterns were integrated using the general structure analysis system (GSAS-II) software [25]. Phase identification was performed using the PDF-5+ database [26], and quantitative phase analysis was carried out by Rietveld refinement with a pseudo-Voigt function using the MDI JADE 10 software package. Microstructural characterization of the samples before and after reduction was conducted using a ZEISS Sigma 500 high-resolution scanning electron microscope (SEM) equipped with an EDAX APEX Advanced X-ray dispersive spectroscopy (EDS) detector, at an acceleration voltage and beam current of 15 kV and 7.7 nA, respectively.

The phase evolution during HyDR for the mixed powder and pre-sintered samples is shown in **Figure 2**. Prior to HyDR, the mixed powder sample exhibits diffraction patterns corresponding to individual metal oxides ($Fe_2O_3$, $Co_3O_4$, $NiO$, and $Mn_2O_3$), as shown in **Figure 2a**. Upon heating to 250–420 °C, the reduction proceeds sequentially through transient intermediate phases, including $Mn_3O_4$ (tetragonal spinel), spinel, and halite. With further heating to 700 °C, the intermediate phases are reduced to FCC, BCC, and MnO. The observed formation of transient spinel and halite phases followed by their reduction to metallic and MnO phases agrees with the thermodynamic prediction [15].

In contrast, the pre-sintered sample initially consists of 55.3±2.1 wt.% spinel and 44.7±1.6 wt.% halite, as shown in **Figure 2d**. No reduction took place in the pre-sintered sample (initially comprised of halite and spinel) until 400 °C. Upon further heating to 600 °C, the spinel phase reduces to a Mn,Fe-rich halite, where metallic Fe (as well as Co and Ni) partitions out of the MnO matrix and form metallic solid solutions, as confirmed by the MnO peaks shifting to lower 2θ and concurrent emergence of BCC and FCC phases (**Figure 2b**). The negligible shift in the mixed powder (4.60° to 4.59°) between 400 °C to 700 °C results from lattice expansion, while in the pre-sintered sample, the order of magnitude larger shift (from 5.42º to 5.30º) reflects reduction and elemental partitioning from the halite phase expanding the lattice [27].

A similar reduction pathway is observed for the Co,Ni-rich halite phase, which completely reduces to FCC phases. Notably, the halite phase is stable until ~600 °C and has relatively higher thermodynamic stability compared with their individual oxide counterparts, *i.e.*, $Co_3O_4$ and NiO, as previously shown by the multicomponent Ellingham diagram [15]. Moreover, the *in-situ* HEXRD data revealed two FCC variants, designated as FCC I ($2\theta_{111}$ = 5.69°) and FCC II ($2\theta_{111}$ = 5.76°). This is due to the difference in elemental concentration of Fe, Co, and Ni in these FCC phases. Based on their lattice parameters (a), it can be concluded that FCC I (a = 3.613 Å) and FCC II (a = 3.574 Å) are Fe-rich and Fe-deficient phases, respectively, due to the larger metallic atomic radius of Fe compared with Co and Ni [28]. Additionally, the onset temperature of BCC phase formation (~470 °C) is lower than that of the FCC I (~560 °C) and FCC II (~640 °C). Since the BCC and FCC phase primarily originates from Fe-rich, and Co, Ni-rich oxides, indicating that the intrinsic stability of chemically mixed multicomponent oxides does not follow the thermodynamic stability trend predicted by the Ellingham diagram [29].

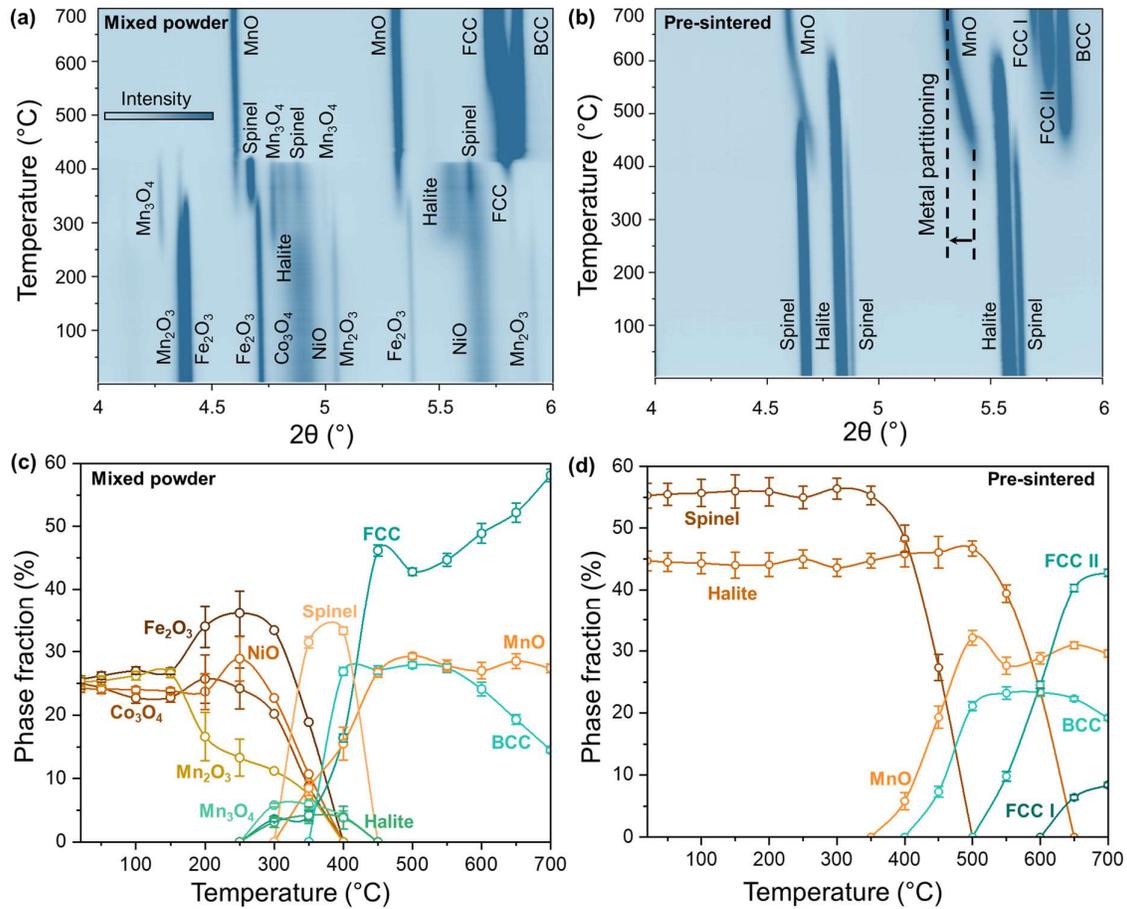

**Figure 2.** Contour maps of the *in-situ* synchrotron high-energy X-ray diffraction (HEXRD) peak intensity as a function of temperature during heating in a hydrogen atmosphere to 700 °C with a ramping rate of 10 °C/min for (a) mixed powder and (b) pre-sintered samples; (c) and (d) the fraction of individual phases for (a) and (b) obtained by Rietveld refinement, respectively.

The phase fractions estimated during HyDR of both samples are shown in **Figure 2c** and **2d**. For the mixed oxide prior to HyDR, the individual metal oxides exhibit phase fractions of 25±0.9 wt.%, confirming the targeted elemental concentration in the initial precursors (**Figure 2c**). As reduction progresses with increasing temperature up to 400 °C, the oxides intermix and undergo reactive sintering, resulting in the formation of 33.3±0.6 wt.% spinel, 4.1±0.8 wt.% $Mn_3O_4$, and 3.8±1.8 wt.% halite. These intermediate oxides ultimately reduce to a mixture of 58.1±1.0 wt.% FCC, 27.4±0.7 wt.% MnO, and 14.5±0.4 wt.% BCC. In contrast, the pre-sintered sample initially consists of 55.3±2.0 wt.% spinel and 44.7±1.6 wt.% halite, as shown in **Figure 2d**. Upon reduction at 700 °C, these solid solution oxides yield 29.6±0.6 wt.% MnO, 19.2±0.4 wt.% BCC, 8.4±0.4 wt.% FCC I (Fe-rich), and 42.8±0.6 wt.% FCC II (Fe-deficient) phases. Although the mixed oxide powders react and form intermediate spinel and halite phases during reduction (~300–450 °C), the sequential reduction steps and phase transformation kinetics are different as is the final phase composition. Notably, the onset of MnO and BCC formation occurs at similar temperatures in the mixed powder (~300–350 °C) and pre-sintered material (~350–400 °C). The minor differences can be attributed to the reduction kinetics of powder versus bulk material, which are even lower for small specimens reduced in a capillary cell compared with larger samples reduced in conventional setups [15]. Nevertheless, FCC metallic phase forms at a significantly lower temperature in the mixed powder (~250 °C) while at ~400 °C for the pre-sintered sample. This correlates with the enhanced thermodynamic stability of the Co,Ni-rich halite phase in the pre-sintered sample. These results demonstrate that the initial precursor critically influences both the reduction pathways and the final phases, owing to the distinct phase composition and altered thermodynamic stability upon sintering prior to HyDR [15].

Microstructures of mixed powder and pre-sintered samples before and after HyDR are presented in **Figure 3**. The mixed powder sample exhibits a porous compact particulate morphology (**Figure 3a** and **3b**) with a homogeneous distribution of fine Co-, Fe-, and Ni-oxide particles (~0.05-1 µm), and larger Mn oxide particles (~0.2-6 µm). Conversely, the initial microstructure of the pre-sintered sample is a dense ceramic with a dual-phase morphology, consisting of Mn,Fe-rich spinel and Co,Ni-rich halite (**Figure 3c** and **3d**). After HyDR, the post reduction powder sample retained a cylindrical shape due to the prior compaction of the powders within the circular capillary cell (**Figure 3e**), while the pre-sintered sample retained its rectangular shape (**Figure 3i**). The morphology of the reduced mixed powder resembles that of the initial mixed powder sample, showing homogeneously distributed partially sintered agglomerates (**Figure 3f-h**). In contrast, the reduced pre-sintered sample exhibits a microstructure comprising two distinct morphologies closely resembling the initial dual-phase microstructure (**Figure 3j** and **3k**). Such two morphological types reflect substantial differences in microstructure evolution of the spinel and halite phases (**Figure 3i-l**). The reduced halite region is relatively dense (~94%) with larger isolated pores (~200 nm), whereas the reduced spinel region is highly porous (~20%) and contains finer pores (~50 nm), owing to the presence of unreduced MnO which inhibits sintering [30].

High magnification SEM micrographs reveal the formation of metallic nanoparticles (~10-50 nm diameter) dispersed throughout the oxide matrix (**Figure 3l**), a phenomenon widely referred to as "exsolution" [31]. In this process, a less stable metal oxide (here, Fe oxide) is initially in solid solution with a relatively highly stable metal oxide (here, Mn oxide). Under reducing atmosphere, the less stable metal oxide metallizes by partitioning out of the oxide matrix and reduces to metal on the surface of the stable oxide. The reasons behind the microstructural evolution lie in the combination of reducing small specimens (~0.3-0.4 mm) and the reducing conditions (e.g., a high local hydrogen partial pressure) in the present case. This feature introduces the possibility of using it as a supported catalyst for heterogenous catalysis, where these metallic nanoparticles (< 20 nm) act as active sites [32]. Additionally, the nanoporous oxide support can enhance catalytic activity by providing a higher surface area, improved mass transport, and increased accessibility of reactants to the active sites [33].

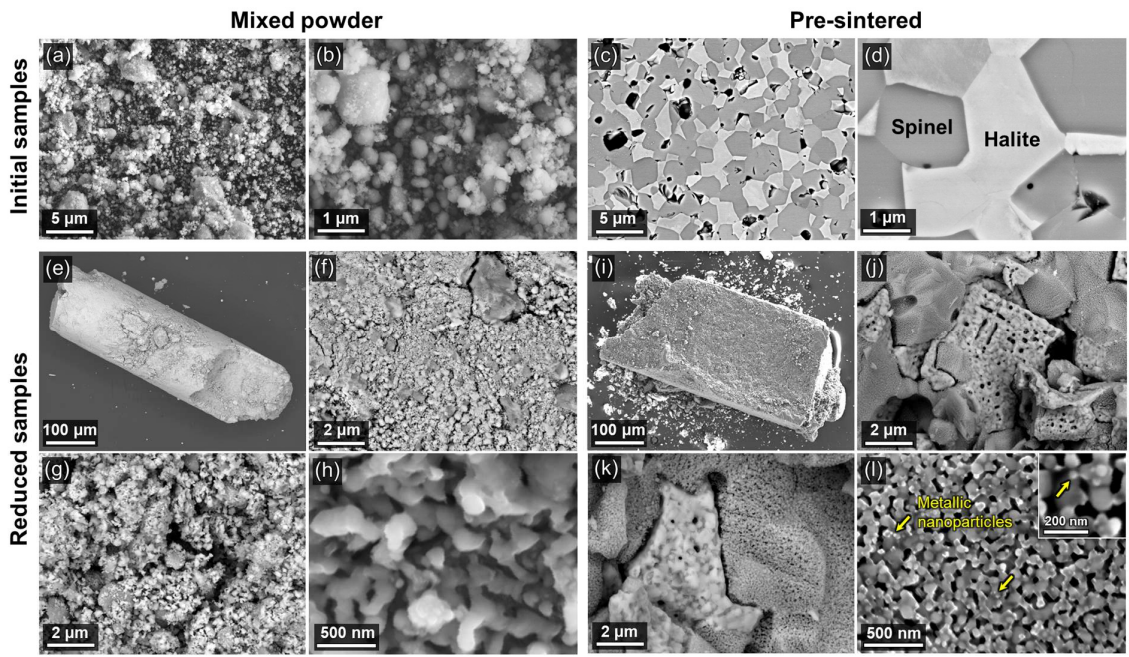

**Figure 3.** SEM micrographs of the precursors before reduction: (a, b) Mixed powder and (c, d) pre-sintered samples. SEM micrographs of the (e-h) mixed powder and (i-l) pre-sintered samples reduced in a hydrogen atmosphere at 700 °C, with a heating rate of 10 °C/min. The arrows point to some of the metallic nanoparticles formed on the surface of a nanoporous oxide matrix.

EDS analysis of the reduced mixed powder and pre-sintered samples is presented in **Figure 4**. The elemental maps for the mixed powder reveal that the Mn signal corresponds to oxygen, confirming the presence of unreduced MnO (**Figure 4b** and **4d)**, along with the formation of an Fe, Co, and Ni-rich metallic phase, as shown by (**Figure 4c, 4e,** and **4f)**. Similarly, the pre-sintered sample shows formation of metallic Fe, Co, Ni and unreduced Mn oxide phases (**Figure 4h-l**). Particularly, the denser regions originated from the Co,Ni-rich halite was reduced to a metallic alloy consisting predominantly of Co and Ni with some Fe and embedded MnO. Whereas, the other regions originating from reduction of the Mn,Fe-rich spinel, involving the exsolution of metal from the stable MnO matrix and resulting in a nanoporous oxide skeleton peppered with abundant metallic nanoparticles. This transformation is evident from the substantial MnO peak shift above 400 °C during the *in-situ* phase evolution of the pre-sintered sample (**Figure 2b**).

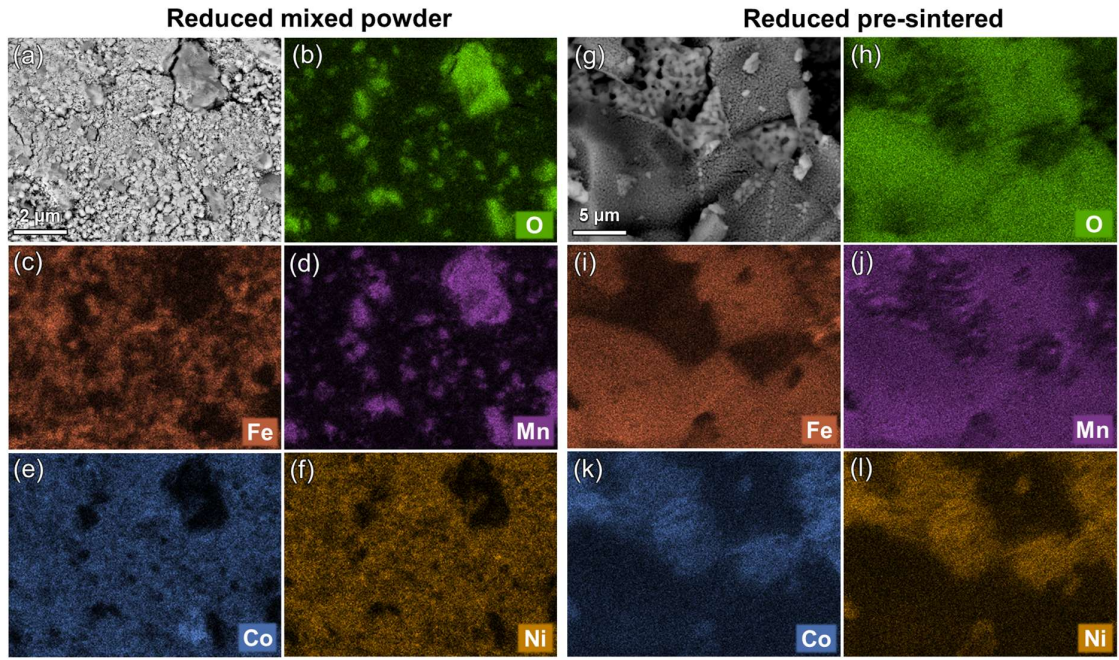

**Figure 4**. (a) SEM micrograph of the reduced mixed powder and corresponding EDS elemental maps of (b) O, (c) Fe, (d) Mn, (e) Co, and (f) Ni. (g) SEM micrograph of the reduced pre-sintered sample and corresponding EDS elemental maps of (h) O, (i) Fe, (j) Mn, (k) Co, and (l) Ni.

Local EDS analysis was further performed to better understand the elemental intermixing and partitioning that took place during the HyDR of pre-sintered sample. EDS point scans were performed within the originally spinel 'grain interior' (spot 1) and at the 'grain boundary' (spot 2), as shown by the marked regions in **Figure 5a**. It was found that although the regions consist mostly of Fe and MnO, there are also considerable amounts of Co and Ni as well (see the EDS spectra and elemental fractions in **Figure 5b**). The metallic phase in the 'grain interior' consists of roughly 35 at.% Fe, 12 at.% Co and 8 at.% Ni. Notably, the 'grain boundaries' are even more enriched with Co and Ni on the expense of MnO. This result suggests that during HyDR process there was segregation of Co and Ni to the spinel grain boundaries, where a larger fraction of them was exsolved. The presence of Co and Ni in the formerly spinel regions (and Fe and MnO in the halite regions) indicates substantial interdiffusion between the different solid solution phases during hydrogen reduction.

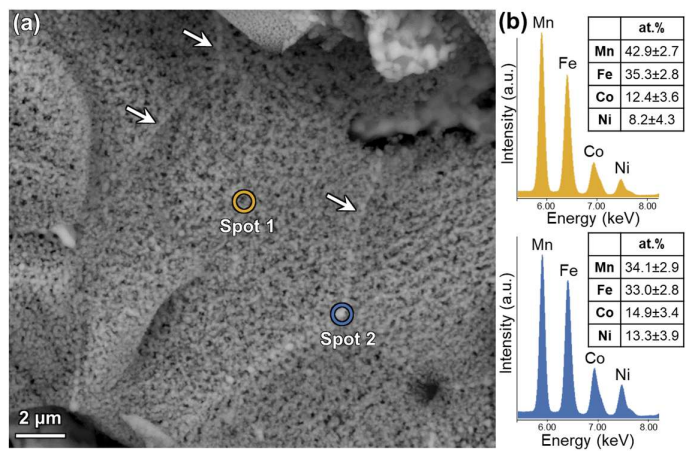

**Figure 5.** (a) SEM micrograph of Mn,Fe-rich region with spot 1 and spot 2 indicating the regions of formerly Mn,Fe-rich spinel 'grain interior' and 'grain boundary' examined by EDS point analysis, some 'grain boundaries' are marked with white arrows. (b) The measured Mn, Fe, Co, Ni transition metals EDS spectra and quantitative results for spot 1 and spot 2, highlighting Co and Ni enrichment decorating at which used to be the spinel grain boundaries.

In summary, hydrogen-based direct reduction (HyDR) of mechanically mixed powder and chemically mixed pre-sintered multicomponent oxide mixtures was investigated via *in-situ* high-energy X-ray diffraction (HEXRD) at 700 ºC. This *in-situ* HEXRD enabled direct tracking of the real time phase evolution during HyDR, thereby providing mechanistic insight into interdiffusion and reduction pathways in multicomponent oxide systems. Both precursor types were reduced to a mixture of metallic and oxide phases yet via distinct reduction routes. The mixed powder was reduced to faced-centered-cubic (FCC), body-centered-cubic (BCC), and MnO through the formation of transient halite, spinel, and $Mn_3O_4$ phases, whereas the pre-sintered sample was directly reduced to BCC, MnO, and FCC (Fe-rich and Fe-deficient) phases. Scanning electron microscopy revealed distinctively different microstructure upon hydrogen reduction of two precursor types. The mixed powder sample showed loosely packed morphology, while the pre-sintered sample developed a porous microstructure comprising Fe-rich metallic particles supported on a MnO matrix. HyDR of multicomponent oxide may offer tunable functionalities (e.g., heterogenous catalysis). These can be further tailored by optimizing precursor selection, particle size, and elemental ratios, as well as by tuning phase evolution during HyDR via controlling hydrogen partial pressure and reducing temperature [34]. Thus, HyDR of multicomponent oxide demonstrates a promising pathway toward designing sustainable functional alloys in a single-step process.

**CRediT authorship contribution statement**

**S. S.** Conceptualization, Methodology, Investigation, Formal analysis, Validation, Visualization, Writing – original draft; **B. R**. Formal analysis, Visualization, Writing – review & editing; **C.P.** – Resources, Writing – review & editing; **D. R**. Supervision, Funding acquisition, Resources, Writing – review & editing; **Y. M.** Conceptualization, Methodology, Supervision, Funding acquisition, Writing – review & editing.

**Declaration of competing interest**

The authors declare that they have no known competing financial interest or personal relationships that could have appeared to influence the work reported in this paper.

**Acknowledgments**

S.S. and Y.M. acknowledges the financial support from Horizon Europe project HAlMan co-funded by the European Union grant agreement (ID 101091936). B.R. is grateful for the financial support of a Minerva Stiftung Fellowship and Alexander von Humboldt Fellowship (Hosted by D.R.). D.R. acknowledges the financial support from the European Union through the ERC Advanced grant ROC (Grant Agreement No. 101054368). Views and opinions expressed are however those of the author(s) only and do not necessarily reflect those of the European Union or the ERC. Neither the European Union nor the granting authority can be held responsible for them. We acknowledge DESY (Hamburg, Germany), a member of the Helmholtz Association HGF, for the provision of experimental facilities. Parts of this research were carried out at P02.1, and we would like to thank Alexander Schökel for his assistance during HEXRD experiment (Proposal I-20231121). We also gratefully acknowledge Jürgen Wichert for his assistance with the sintering and specimen preparation and Rebecca Renz for the scientific illustration, both at the Max Planck Institute for Sustainable Materials.